\documentclass[fleqn]{2017SCGE}

\usepackage{natbib}
\setcitestyle{numbers,square}
\usepackage{float}
\usepackage{amsmath}
\DeclareMathOperator{\dprime}{\prime \prime}
\DeclareMathOperator{\arcsec}{\prime \prime}
\usepackage{color,tabularx}
\usepackage{graphicx}

\makeatletter
\let\jnl@style=\rmfamily 
\def\ref@jnl#1{{\jnl@style#1}}%
\newcommand\aj{\ref@jnl{AJ}}%
\newcommand\psj{\ref@jnl{PSJ}}%
\newcommand\araa{\ref@jnl{ARA\&A}}%
\newcommand\apj{\ref@jnl{ApJ}}%
\newcommand\apjl{\ref@jnl{ApJL}}     %
\newcommand\apjs{\ref@jnl{ApJS}}%
\newcommand\ao{\ref@jnl{ApOpt}}%
\newcommand\apss{\ref@jnl{Ap\&SS}}%
\newcommand\aap{\ref@jnl{A\&A}}%
\newcommand\aapr{\ref@jnl{A\&A~Rv}}%
\newcommand\aaps{\ref@jnl{A\&AS}}%
\newcommand\azh{\ref@jnl{AZh}}%
\newcommand\baas{\ref@jnl{BAAS}}%
\newcommand\icarus{\ref@jnl{Icarus}}%
\newcommand\jaavso{\ref@jnl{JAAVSO}}  %
\newcommand\jrasc{\ref@jnl{JRASC}}%
\newcommand\memras{\ref@jnl{MmRAS}}%
\newcommand\mnras{\ref@jnl{MNRAS}}%
\newcommand\pra{\ref@jnl{PhRvA}}%
\newcommand\prb{\ref@jnl{PhRvB}}%
\newcommand\prc{\ref@jnl{PhRvC}}%
\newcommand\prd{\ref@jnl{PhRvD}}%
\newcommand\pre{\ref@jnl{PhRvE}}%
\newcommand\prl{\ref@jnl{PhRvL}}%
\newcommand\pasp{\ref@jnl{PASP}}%
\newcommand\pasj{\ref@jnl{PASJ}}%
\newcommand\qjras{\ref@jnl{QJRAS}}%
\newcommand\skytel{\ref@jnl{S\&T}}%
\newcommand\solphys{\ref@jnl{SoPh}}%
\newcommand\sovast{\ref@jnl{Soviet~Ast.}}%
\newcommand\ssr{\ref@jnl{SSRv}}%
\newcommand\zap{\ref@jnl{ZA}}%
\newcommand\nat{\ref@jnl{Nature}}%
\newcommand\iaucirc{\ref@jnl{IAUC}}%
\newcommand\aplett{\ref@jnl{Astrophys.~Lett.}}%
\newcommand\apspr{\ref@jnl{Astrophys.~Space~Phys.~Res.}}%
\newcommand\bain{\ref@jnl{BAN}}%
\newcommand\fcp{\ref@jnl{FCPh}}%
\newcommand\gca{\ref@jnl{GeoCoA}}%
\newcommand\grl{\ref@jnl{Geophys.~Res.~Lett.}}%
\newcommand\jcp{\ref@jnl{JChPh}}%
\newcommand\jgr{\ref@jnl{J.~Geophys.~Res.}}%
\newcommand\jqsrt{\ref@jnl{JQSRT}}%
\newcommand\memsai{\ref@jnl{MmSAI}}%
\newcommand\nphysa{\ref@jnl{NuPhA}}%
\newcommand\physrep{\ref@jnl{PhR}}%
\newcommand\physscr{\ref@jnl{PhyS}}%
\newcommand\planss{\ref@jnl{Planet.~Space~Sci.}}%
\newcommand\procspie{\ref@jnl{Proc.~SPIE}}%

\newcommand\actaa{\ref@jnl{AcA}}%
\newcommand\caa{\ref@jnl{ChA\&A}}%
\newcommand\cjaa{\ref@jnl{ChJA\&A}}%
\newcommand\jcap{\ref@jnl{JCAP}}%
\newcommand\na{\ref@jnl{NewA}}%
\newcommand\nar{\ref@jnl{NewAR}}%
\newcommand\pasa{\ref@jnl{PASA}}%
\newcommand\rmxaa{\ref@jnl{RMxAA}}%

\newcommand\maps{\ref@jnl{M\&PS}}%
\newcommand\aas{\ref@jnl{AAS Meeting Abstracts}}%
\newcommand\dps{\ref@jnl{AAS/DPS Meeting Abstracts}}%
\makeatother

\newcommand{\checkit}[1]{#1}

\newcommand{\msun}{M_\odot}

\newcommand{\mh}{{\rm [M/H]}}
\newcommand{\gaia}{\textit{Gaia~}}
\newcommand{\trilegal}{\texttt{TRILEGAL}}
\newcommand{\phoenix}{\texttt{PHOENIX}}
\newcommand{\CKmodel}{\texttt{CK04}}
\newcommand{\parsec}{\texttt{PARSEC}}
\newcommand{\healpix}{\texttt{HEALpix}}
\newcommand{\healpy}{\texttt{healpy}}

\newcommand{\nside}{n_\text{side}}
\newcommand{\npix}{n_\text{pix}}
\newcommand{\ebv}{E(B-V)}
\newcommand{\um}{\mu m}

\makeatletter
\def\vector(#1){%
\mathchoice%
    {\pmatrix@i{\vector@i#1,,}}%
    {\pmatrix@ii{\vector@i#1,,}}%
    {}%
    {}%
}
\def\vector@i#1,{\if,#1,\else{#1}\cr\expandafter\vector@i\fi}
\def\pmatrix@i#1{\begin{pmatrix}#1\end{pmatrix}}
\def\pmatrix@ii#1{\left(\!\begin{smallmatrix}#1\end{smallmatrix}\!\right)}
\makeatother

\begin{document}

\ensubject{subject}

\ArticleType{Article}%
\Year{2023}
\Month{June}
\Vol{00}
\No{0}
\ArtNo{000000}
\ReceiveDate{June 10, 2023}

\title{The first comprehensive Milky Way stellar mock catalogue for the Chinese Space Station Telescope Survey Camera}{CSST MW stellar mock catalogue}

\author[1,2]{Yang Chen}{cy@ahu.edu.cn}%
\author[3]{Xiaoting Fu}{xiaoting.fu@pmo.ac.cn}
\author[2]{Chao Liu}{}%
\author[4]{Piero~Dal~Tio}{}
\author[5]{L\'eo Girardi}{}
\author[5]{\\Giada Pastorelli}{}
\author[4]{Alessandro Mazzi}{}
\author[4]{Michele Trabucchi}{}
\author[2]{Hao Tian}{}
\author[2,6]{Dongwei Fan}{}
\author[4]{\\Paola Marigo}{}
\author[7]{Alessandro Bressan}{}

\AuthorMark{Chen Yang}
\AuthorCitation{Chen Y. et al}

\address[1]{School of Physics and optoelectronic engineering, Anhui University, Hefei 230601, China}
\address[2]{National Astronomical Observatories, Chinese Academy of Sciences, Beijing 100101, China}
\address[3]{Purple Mountain Observatory, Chinese Academy of Sciences, Nanjing 210023, China}
\address[4]{Dipartimento di Fisica e Astronomia Galileo Galilei, Universit\`a di Padova, Vicolo dell'Osservatorio 3, I-35122 Padova, Italy}
\address[5]{INAF - Osservatorio Astronomico di Padova, Vicolo dell'Osservatorio 5, I-35122 Padova, Italy}
\address[6]{National Astronomical Data Center, Beijing 100101, China}
\address[7]{SISSA, via Bonomea 365, I-34136 Trieste, Italy}

\abstract{
The Chinese Space Station Telescope (CSST) is a cutting-edge two-meter astronomical space telescope currently under construction. Its primary Survey Camera (SC) is designed to conduct large-area imaging sky surveys using a sophisticated seven-band photometric system. The resulting data will provide unprecedented data for studying the structure and stellar populations of the Milky Way. 
To support the CSST development and scientific projects related to its survey data, we generate the first comprehensive Milky Way stellar mock catalogue for the CSST SC photometric system using the \trilegal\, stellar population synthesis tool. The catalogue includes approximately 12.6 billion stars, covering a wide range of stellar parameters, photometry, astrometry, and kinematics, with magnitude reaching down to $g$=27.5 mag in the AB magnitude system. 
The catalogue represents our benchmark understanding of the stellar populations in the Milky Way, enabling a direct comparison with the \checkit{future} CSST survey data. 
Particularly, it sheds light on faint stars that are hidden from current sky surveys. 
Our crowding limit analysis based on this catalogue provides compelling evidence for the extension of the \checkit{CSST Optical Survey (OS)}  to cover low Galactic latitude regions. 
The strategic extension of the \checkit{CSST-OS} coverage, combined with this comprehensive mock catalogue, will enable transformative science with the CSST.%
}


\maketitle

\begin{multicols}{2}
\section{Introduction}\label{sec:intro}

The Chinese Space Station Telescope (CSST) is a two-meter astronomical space telescope that will be sent to\Authorfootnote 
dock with the Chinese Space Station (CSS, also known as Tiangong). The first core module of the space station (Tianhe; TH) was launched in April 2021. Approved in 2013, the telescope is expected to launch in the near future
 and will be freely orbiting in the same path as the CSS. It is designed to be serviceable  (for human manipulations) while docked with the CSS \citep{Su14,Zhan18}. %
 The science goals of CSST include exploring a range of important questions in cosmology, galaxies, stars, and exoplanets. To achieve these goals, it will be equipped with several astronomical instruments including the Survey Camera (SC), Terahertz Receiver, Multichannel Imager, Integral Field Spectrograph, and Cool-Planet Imaging Coronagraph.
The SC, in particular, will carry out a seven-year-long wide-area multi-band imaging and slitless spectroscopic survey, as well as other key programs and Guest Observer (GO) programs for an additional two years \cite{Zhan18}.

The Chinese Space Station Optical Survey (CSS-OS) will utilize the SC to take images and obtain slitless spectra of a vast 17,500 deg$^2$ sky area for Galactic latitude $b>15^\text{o}$  \cite{Zhan18}. The depth for the imaging survey is expected to reach $\sim26.3$\,mag in the $g$ band ($5\sigma$ for point sources in AB magnitude system). Additionally, a smaller 400 deg$^2$ area will be surveyed to a depth of $\sim1$ magnitude deeper. 
This latter ultra-deep survey will cover well-known fields that have been previously cataloged by other surveys from X-ray to sub-millimeter, including COSMOS, XMM-LSS, GOODS-N, and ECDFS \cite{XMM-LSS, COSMOS, GOODS-N,ECDFS}.

The CSS-OS ensures a large sky coverage through CSST's wide field of view, approximately $1.1^\circ \times 1.0^\circ$. Additionally, the telescope offers high quality \checkit{data} due to its exceptional spatial resolution, characterized by a full width at half maximum (FWHM) of the point spread function (PSF) at approximately $0.15\arcsec$, and a pixel scale of 0.074$\arcsec$ \cite{Zhan21}. It provides high photometric precision, with a standard deviation of approximately 0.01 mag at $g \sim 20$ mag. 
Though these image quality capabilities are comparable to those of the UVIS channel of the Hubble Space Telescope's wide field camera 3 (WFC3) \cite{wfc3}, the CSST's advantage lies in its field of view, which is over 500 times larger, making it an unprecedented ideal instrument for efficient sky surveys.

In addition to addressing questions related to dark matter, dark energy, and other cosmological topics  \cite{Cao18, Gong19, Zhang19, Zhou21}, the survey data collected by CSST will be crucial in understanding the structure and stellar population properties of the Milky Way (MW) and nearby galaxies, thanks to its large sky coverage and high quality.
The MW is a complex and dynamic system comprising of multiple main components such as thin and thick disks, bulge and halo, as well as sub-structures including star clusters, streams and overdensities). To accurately map the Galaxy and reveal its properties in detail, it is essential to have high-quality imaging surveys with wide sky coverage. 
A notable example highlighting the significance of such surveys is the discovery of  \gaia Enceladus \cite{Helmi2018}, the structure of the major merger event in the Milky Way's history. This discovery was made possible  by the \gaia mission's full-sky, high-precision observations.

The project ``Studying the Stellar Population of Milky Way and Nearby Galaxies''  has been selected as one of the first scientific preparation initiatives for CSST (CMS-CSST-2021-A08). The project aims to achieve \checkit{several} key objectives, including:
$1)$ Investigating the stellar properties of  Galactic stars, Galactic clusters, and stars in nearby galaxies using mock catalogues;
$2)$ Developing algorithms for determining stellar parameters and identifying anomalous stars using photometric and slitless spectra data;
$3)$ Identifying M stars and brown dwarfs with the use of advanced algorithms;
$4)$ Evaluating methods for accurately determining the stellar initial mass function of low-mass stars
and calibrating stellar parameters based on open cluster member stars;
$5)$  Examining star clusters in the MW and nearby galaxies; 
and $6)$  Conducting research on stellar populations, star clusters, stellar streams, and stellar dynamics.

Mock stellar catalogues are essential tools for these project.
They serve multiple important functions, including calibrating the photometric system's quality, validating data reduction pipelines, informing survey strategy, and enabling preparation-stage research on Milky Way structures, low-mass stars, initial mass functions, and star clusters. Additionally, mock catalogues provide valuable information on the Milky Way foreground sources, which is crucial for extragalactic studies.
One notable example of a successful mock stellar catalogue is the \gaia Universe model \cite{Robin2012}. This catalogue informed and optimized the observing strategy during the development stage of the \gaia mission \cite{Gaia2016} and serves as a benchmark for real \gaia data analysis. For instance, it helped to refresh our view of the Milky Way's formation history \cite{Helmi2018} and analyze stellar populations with different properties \cite{Cantat-Gaudin2020}. 
By using mock data as a proxy for future CSST survey data, the project will be able to conduct studies
on the structure, properties, and dynamics of the Milky Way and nearby galaxies

To realize the above potential science applications for CSST, we present a MW stellar mock catalogue for the CSST SC photometric system. 
\checkit{Our simulation tools and strategies are based on those used in our work for the Large Synoptic Survey Telescope (LSST) as described by Dal Tio et al. (2022) \cite{DalTio2022}, with some modifications to account for the CSST instrument.}
While the LSST simulation  \cite{DalTio2022} focused mainly on the southern sky due to the limitation of ground-based observations, our MW stellar mock catalogue simulates the full sky for the space telescope CSST.    
Another difference concerns the treatment of binaries. The specific science goals of LSST required Dal Tio et a. (2022) \cite{DalTio2022} to put some care into describing the properties of interacting binaries in detail, whereas we concentrate on simulating photometric binaries  for this release.
However, we plan to incorporate more detailed simulations of interacting binaries in future releases.

This work is structured as follows. We introduce the CSST SC photometric system in section \ref{sec:csst_photo} and describe our method and the codes we used in section \ref{sec:method}. In sections \ref{sec:sim} and \ref{sec:cluster}, we present the results of our simulation for the Galactic components and star clusters, respectively. In section \ref{sec:sum}, we summarize the simulation results.

\section{CSST photometric system} \label{sec:csst_photo}

\begin{figure*}%
\centering
\includegraphics[scale=0.55]{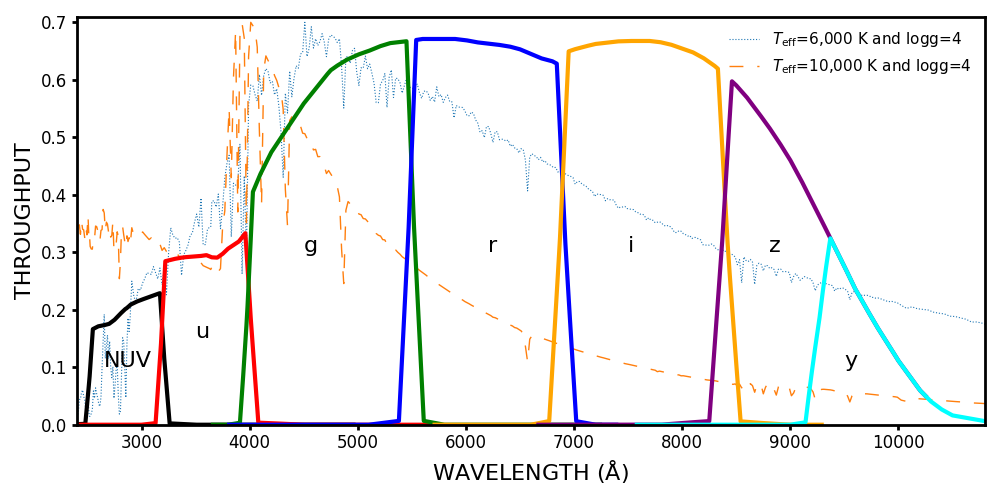}
\caption{CSST filter transmission curves (solid curves with their corresponding band names). For comparison, two scaled stellar SEDs, with solar metallicity from Castelli \& Kurucz  \cite{ck03}, are also displayed.}
\label{fig:filter}
\end{figure*}

The CSST SC employs a set of seven filters covering a wide wavelength range from 2000\,\AA~ to 1.1\,$\um$, which includes the ${\rm NUV}$, $u$, $g$, $r$, $i$, $z$, and $y$ bands. 
Figure \ref{fig:filter} displays the CSST filter transmission curves for each of the filters, indicated by solid curves with their corresponding band names. Additionally, the figure includes two scaled stellar spectral energy distributions (SEDs) of two star with different \checkit{effective} temperature.
This set of filters enable sampling of several key stellar features, including the Balmer break at $\sim$ 3646\,\AA, the D$_{4000}$ break at $\sim$ 4000\,\AA\, \cite{break4000}, and the Rayleigh-Jeans slope, as well as the flux peak for most stars. 
With these features, the filters are well-suited for deriving stellar parameters through multi-band SED fitting, color-magnitude diagram (CMD), and color-color diagrams for a large sample of stars with a wide range of physical properties.

To contextualize CSST within the landscape of forthcoming wide-field optical space telescopes, it is illustrative to compare its specifications with those of the Euclid  \cite{Euclid} and Nancy Grace Roman Space Telescope (Roman, formerly WFIRST)  \cite{WFIRST}. Table~\ref{tab:telescope} of the appendix presents a comparison among the imaging instruments of the three telescopes. The CSST has a larger field of view (FoV) and bluer filters, providing advantages over Euclid and Roman, which have filters covering longer wavelengths. While Roman has a smaller FoV and will survey a smaller area, it will reach much fainter magnitudes. Therefore, the survey data collected by these telescopes will be complementary, and they will mutually expand the research possibilities offered by each other.
In addition to these telescopes, it is impossible to ignore the critical contribution of the \gaia mission  \cite{Gaia2016}, which has produced unprecedented data by repeatedly scanning the sky. 
Yet, the \gaia survey does not reach especially faint magnitudes as the CSST will do, therefore a synergy between the two can be expected, for instance in terms of consolidating astrometric data as proposed by Gai et al. (2022) \cite{Gai22}.

\section{Method and codes: TRILEGAL} \label{sec:method}

The main tool we use to generate the stellar mock catalogue is \trilegal, a stellar population synthesis code that uses a star counting approach \cite{Girardi05,Girardi12,Girardi16}. \checkit{{\trilegal} can} simulate both deep and/or wide-area photometric surveys for any given photometric systems. This flexibility makes it a powerful tool for generating realistic synthetic populations of stars for a wide range of scientific investigations.

\trilegal~ utilizes a process of interpolating among a set of stellar evolutionary tracks to generate isochrones. 
Though it has been tested primarily with the \parsec\ stellar model database \cite{parsec},
 it can also accept other stellar track databases as input. Through a Monte Carlo approach, it randomly samples these isochrones based on specified stellar distribution functions. These functions encompass: 
 1) the initial mass function (IMF), which describes the distribution of stellar masses within a given total mass;
 2) the star formation history (SFH), which indicates the number of stars formed during specific time periods; 
and 3) the age-metallicity relation (AMR), which defines the metallicity distribution of stars for a given age.
These distribution functions are normalized to a given total stellar mass. 

In the context of MW simulations, the total stellar mass at a given position is determined using the MW geometric model, which incorporates specific stellar components such as the Galactic thin disk, thick disk, halo, and bulge. 
The calibration of these stellar components relies on observations using number counts and luminosity functions (see Sec. \ref{subsec:compo} for details). When simulating star clusters, the total stellar mass is calibrated based on the target star cluster.
Each star generated randomly from the distribution functions is then interpolated among the isochrones to calculate its photometric properties using the embedded YBC code \cite{Chen19}. Variabilities of the stars are determined based on pulsation models \cite{Trabucchi2019,Trabucchi2021}, and their kinematics are derived from empirical relations  \cite{Chiba00,Robin03,Holmberg09}. 
Regarding photometric properties, Galactic extinction is computed using the Planck MW dust map \cite{Planck-dust}. 
Finally, \trilegal~outputs various information for each simulated star, including basic stellar physical parameters (luminosity, effective temperature, gravity, etc.), photometric data (magnitudes in different filters), kinematics, and variability periods.
In the following sections, we will provide a comprehensive explanation of the key components mentioned earlier in \trilegal~ to generate the MW stellar population.

\subsection{Geometric models of MW stellar components}
\label{subsec:compo}

The MW model used in the current simulation incorporates several stellar components, namely the thin disk, thick disk, halo, and bulge. Each component is characterized by its own three-dimensional mass density distribution, SFH and AMR. However, all of these components share the same IMF as defined by Chabrier et al. (2003)  \cite{Chabrier03}. 
Here we outline the assumptions made for each of these stellar components. These assumptions and model parameters have been derived from previous calibration efforts \cite{Girardi05,Pieres2020}.

\begin{itemize}
\item The thin disk component is characterized by a stellar mass density that follows an exponential function along the radial direction on the Galactic plane and a squared hyperbolic secant $\mathrm{sech}$ function in the vertical direction:
\begin{equation*}
    \rho_{\rm thin}(R_{\rm p}, Z) \propto
    {\rm e}^{-(R_{\rm p}-R_\odot)/h_r}\frac{R_{\rm p}}{R_\odot}\frac{1}{2h_z}\left(e^{-Z/2h_z}+e^{Z/2h_z}\right)^{-2}
\end{equation*}
where 
\begin{equation*}
    h_z=h_{z,0}(1+\frac{t}{5.55\,{\rm Gyr}})^{5/3}
\end{equation*}
following Rana \& Basu (1992)  \cite{Rana&Basu92}  but with different parameters, scale height ($h_{z,0}=$94.69\,pc) and scale radius ($h_r=$2.91\,kpc). ${R_\odot}=8.7$kpc 
 is the solar distance to the Galactic center. $R_{\rm p}$ is the radius projected onto the Galactic plane and is truncated at 25\,kpc. 
Although L{\'o}pez-Corredoira et al. (2018) 
 \cite{Lopez18} have argued the disk truncation should be beyond 25 kpc, we find only minor differences when testing values of the truncation radius at 25, 30, and 35 kpc.
\checkit{The term $\frac{R_{\rm p}}{R_\odot}$} is introduced to moderate the exponential increase towards the Galactic center. In terms of normalization, the local surface density for the thin disk is defined as the total mass of thin disk stars ever formed in the Solar Neighbourhood, integrated vertically over a unit area. It has been calibrated to 55.41\,$\msun {\rm pc}^{-2}$  \cite{Pieres2020}. Additionally, we assume a constant SFR over the last 11 Gyr and the AMR is adopted from Rocha-Pinto et al. (2000) \cite{Rocha-Pinto00}.
\item The thick disk component of the MW model also follows an exponential function in the radial direction and a squared  $\mathrm{sech}$ function in the vertical direction.  However, it has  a fixed scale height $h_z=$800\,pc, a scale length $h_r=$2.39\,kpc, and a local thick disk surface density as 0.0010\,$\msun {\rm pc}^{-2}$ (which represents the vertically integrated total mass of thick disk stars ever formed in the Solar neighbourhood per unit area). The star formation rate is assumed to be constant between 11\,Gyr and 12 Gyr ago, and zero for other time periods. The metallicity is fixed at $Z=0.004$ with a spread of $\sigma({\rm log}Z)=0.15$. 
\item The halo component in the MW halo is represented by an axisymmetric power-law profile: 
\begin{equation*}
\rho_{\rm h} = \rho_{\rm h,\odot}\left(\frac{\sqrt{R_{\rm p}^2+(Z/q)^2}}{R_\odot}\right)^{-\alpha}
\end{equation*}
The parameters are set as $q=0.62$ and $\alpha=2.75$. Here, $\rho_{\rm h,\odot}$ is the total mass of the halo stars ever formed in the Solar neighbourhood per unit volume and is set to 0.0001\,$\msun {\rm pc}^{-3}$. 
To ensure a flat central density, a minimum radius of 0.5 kpc in R=($\sqrt{R_{\rm p}^2+(Z/q)^2}$) is applied.
The SFR in the halo is assumed to be constant over the time 12 to 13 Gyr ago and zero for other time. The AMR adopted is based on Henry \& Worthey (1999)  \cite{Henry99}.
\item The bulge is described by a triaxial system with a truncated power-law as in Binney et al. (1997)  \cite{Binney97}: 
\begin{equation*}
    \rho_{\rm b}=\rho_0\frac{{\rm e}^{-a^2/a_m^2}}{(1+a/a_0)^{1.8}}, 
\end{equation*}
where 
\begin{equation*}
a=\left(x^2+\frac{y^2}{\eta^2}+\frac{z^2}{\zeta^2}\right)^{1/2}
\end{equation*}
with $a_m=$2.5\,kpc, $a_0=$95\,pc, $\eta=0.68$ and $\zeta=0.31$. $\rho_0$ is the total stellar mass ever formed in the Galactic center per unit volume and is set to 406\,$\msun {\rm pc}^{-3}$. The SFH and AMR are from Zoccali et al. (2003)  \cite{Zoccali03}.
\end{itemize}

The calibrated parameters mentioned above can be further refined and improved using existing data, such as \gaia data release. Additionally,  future large area survey data, including those from projects like the LSST and CSST will contribute to the refinement of these parameters and enable more accurate MW modelling.

 In addition to modelling the MW components, \trilegal\, has the capability to simulate optional objects located at fixed distances, such as the LMC and SMC, and Galactic or extra-galactic star clusters. 
 In this work, mock catalogues for star clusters are also \checkit{generated,} while future dedicated projects will focus on \checkit{modelling} the LMC, the SMC and the Andromeda galaxy.

\subsection{Milky Way interstellar dust map}\label{sec:dust}
The interstellar dust is a non-negligible components of Galactic models, especially for the disks and bulge components. The MW dust model employed in this simulation is taken from the two-dimensional $\ebv$ map provided by the Planck collaboration %
 \cite{Planck-dust}. We adopt $\ebv =  \ebv_\mathrm{xgal}$ for $\ebv_\mathrm{xgal}<0.3$~mag and $\ebv = 1.49\times10^4\times\tau_{353}$ otherwise, as recommended by the Planck Collaboration, where $\ebv_\mathrm{xgal}$ and $\tau_{353}$ are the $\ebv$ for extra-galactic studies and the optical depth at 353\,GHz as stored in the Planck dust map respectively. $R_V=3.1$ is then used to convert $\ebv$ values to $A_V$ (=$R_V*\ebv$). 
 The adopted dust map used in this study provides \checkit{a full} sky coverage and has  $\nside=2048$ ({\healpix} pixelization) or a resolution corresponding to 5$\prime$ per pixel. 
 This map represents the total extinction along the line of sight and assumes an exponential distribution of dust as a function of distance from us. While there are some three-dimensional dust maps in the literature that offer a more reasonable distribution of the dust along the line of sight, they are not used in this study due to their limited sky coverage  \cite{Green2019}. 

We compared the Planck mean extinction map with \checkit{the Schlegel one \cite{Schlegel}}, and found a high level of agreement between the two maps, with differences typically within 1 percent for the majority of regions (see appendix \ref{app:dust}). 
Larger differences were observed in dust-shrouded regions and in the direction of external galaxies, such as the Magellanic clouds and the Andromeda galaxy. 
In these specific areas, the Planck dust map predicts lower levels of extinction compared to the Schlegel map. 
Taking into account both resolution and sky coverage, we have chosen to adopt the Planck map as our preferred option.

\subsection{\parsec\, stellar evolutionary tracks}

To simulate the MW stellar population, stellar models are required as an input. The stellar model we currently use in \trilegal\ is \parsec\ V1.2s.
It is based on the solar-scaled evolutionary tracks presented in Bressan et al. (2012)  \cite{parsec}, 
and with important updates to low mass stars  \cite{Chen14}, very massive stars  \cite{Chen15,Tang14} and  TP-AGB stars  \cite{Marigo17,Pastorelli19,Pastorelli20}. 
For modelling non-solar-scaled stellar populations especially at low metallicities, we use $\alpha$-enhanced stellar models from Fu et al. (2018)  \cite{Fu18}. 
These $\alpha$-enhanced models are currently only applied for old metal-poor star clusters.
It is an approximation to apply the non-$\alpha$-enhanced stellar models to the Galactic field stars in the Galactic bulge, thick disk, and halo,
where $\alpha$-enhanced stellar populations are found  \checkit{\cite{Maraston2003,Ishigaki2012,Bensby2014,Rix2022}. We} have started the \parsec\ 2.0 project to include $\alpha$-enhancement  together with rotation for a wide range of stellar mass and metallicity  \cite{Nguyen2022}.
We will include this full set of \parsec\ 2.0 stellar models in \trilegal\ once it is available.

\subsection{YBC bolometric correction database}
Given the theoretical isochrones, we need to convert the physical quantities into observable ones, that is to say the magnitudes or spectral energy distributions. This is done by looking for the bolometric correction (BC) models corresponding to stars of different physical parameters (effective temperature, gravity, metallicity, etc.). In this simulation, we use the YBC code  embedded in \trilegal\, and the corresponding database, which provides BC models for a broad range of stellar parameters. 
\checkit{
In short, YBC combines different atmosphere models in a unified BC database and applies them to stars with diverse parameters.
}
These BC models for the CSST SC photometric system are also provided on the YBC online \checkit{database\footnote{\url{https://sec.center/ybc/}}.} 
The extinction of the interstellar medium is also handled by YBC code with the consideration of the effect brought by the star-by-star spectral variation.

\subsection{Sky pixelization with \healpix}
\trilegal\, is a single-threaded program, which simulates the stars within the cone volume for a given pointing on the sky. Therefore, to simulate the full sky, we need to split the sky into small pieces, and initiate single \trilegal\, simulations for each of them in parallel. For this purpose, we pixelize the full sky with the help of the widely used {\healpix} realization  \cite{healpix}. 
Every small simulation for the corresponding sky pixel represents a conic section along the line of sight through the MW. With its central coordinate and area of a given pixel, the extinction is read from the extinction map, and \trilegal\, computes its stellar content.
The pixel resolution should be high enough so that the field variation of the properties of the stellar populations can be modelled accurately. However, a too high resolution requires a very high amount of computational time and storage, and the pixel may contain too few stars, leading to large stochastic errors. Therefore, an optimal pixelization scheme would be to divide the sky with variable pixel resolutions, which is enabled with the \healpix\, package. With \healpix, the sky is divided hierarchically into $12*\nside^2$ ($\nside=2^0,2^1, 2^2, ..., 2^n, ...$) equal area pixels (though not of the same shape), and each pixel is designated with a unique index ($\npix$). For a given $\nside$, the sky  coordinate of the pixel center for each $\npix$ is determined. Therefore, the desired pixelization scheme is enabled with the combination of several different $\nside$s. 

In this work, we consider variable pixel resolutions of $\nside=64, 128, 256, 512, 1024$ and 2048.
The largest value, $\nside=2048$, is adopted because it is the highest resolution of the Planck dust map we use (see section \ref{sec:dust}). Starting with $\nside=64$, we calculate the variations of the extinction and mass column density, which are based on the quantities at the next higher $\nside$ considered. If the standard deviation ($\sigma_\text{ext}$) of $\ebv$, the relative error ($\sigma_\text{ext}/\ebv$), and the standard deviation of the surface mass density ($\sigma_\rho$) are larger than the thresholds (0.3, 0.1 and 0.1 for $\sigma_\text{ext}/\ebv$, $\sigma_\text{ext}$ and $\sigma_\rho$ respectively), we increase the resolution until these thresholds are reached but no finer than $\nside=2048$.
In the end, we have 596,601 pixels with the resolution $\nside$ ranging from 64 to 2048. Each pixel is labelled with a unique combination of $\nside$ and $\npix$. This step is done with the Python \healpy\, package  \cite{Zonca19}.

\begin{figure*}
\centering
\includegraphics[width=0.65\textwidth]{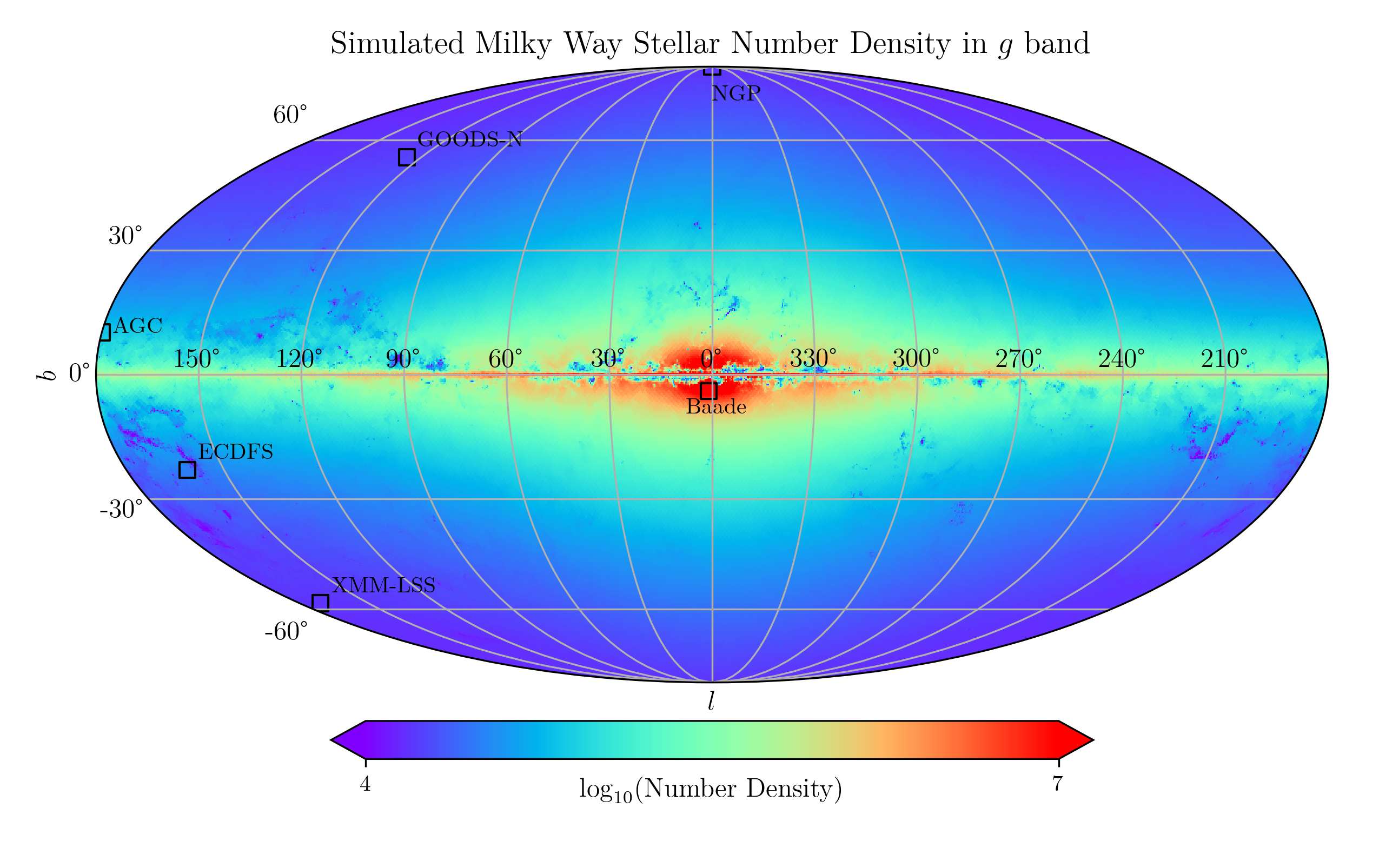}
\caption{Stellar surface number density (per square degrees) in the CSST $g$ band (< 27.5 mag). The resolution is with \healpix\ $\nside=2048$. %
}
\label{fig:density}
\end{figure*}

\section{Simulation results for Galactic components} \label{sec:sim}
In this section, we present our simulation results for the Galactic components. 

\subsection{The mock catalogue}
The mock catalogue contains 596,601 individual fits format files (gz-compressed) corresponding to each of the sky pixels. The total size of the mock catalogue data is about 1.5 Terabytes. It can be requested for express delivery for the whole catalogue, or internet transfer if only a small portion is in need. Figure~\ref{fig:density} gives an overview of the simulated stars in the form of the stellar number density map in the $g$ band. It contains a total number of about 12.6 billion stars down to the magnitude of $g=27.5$ mag. Although according to the current CSS-OS planning, it will reach $g\sim26.3$\,mag for the 17,500 deg$^2$ and $g\sim27.5$\,mag for another 400 deg$^2$, we take the limit $g=27.5$\,mag to give room for adding photometric errors and for exploring possible changes to the survey strategy. From this figure, we also see the non-smooth features caused by the variation in the dust map, since the intrinsic stellar components are modelled by some smooth analytical formulae as described in section \ref{subsec:compo}. We will discuss the dust effects of using different dust maps on the stellar number density distribution in a more detailed study.

We show two rows of an example fits file in table~\ref{tab:cat} of the appendix and the meaning of the columns are described in appendix \ref{app:table_description}. In summary, we provide detailed information on stellar physical parameters, photometry, astrometry and kinematics for each of the simulated stars. We caution that the binary stars in this simulation are non-interacting binaries. The results with the code including the interacting binaries, developed in  Dal Tio et al. (2021)  \cite{DalTio21}, 
will be presented in the next releases.

\subsection{Mock catalogue distribution through the Virtual Observatory}

The full set of the raw catalogue can be accessed and retrieved from the Chinese Virtual Observatory data center\footnote{\url{https://nadc.china-vo.org/data/data/csst-trilegal/f}}.
However, due to resource limitations and constraints within the VO framework, 
the current online version of  the mock catalogue is presented at a uniform pixel resolution of $\nside=128$.
This resolution is deemed sufficient for the majority of applications that use this catalogue.

\begin{figure*}
\centering
\includegraphics[width=0.78\textwidth]{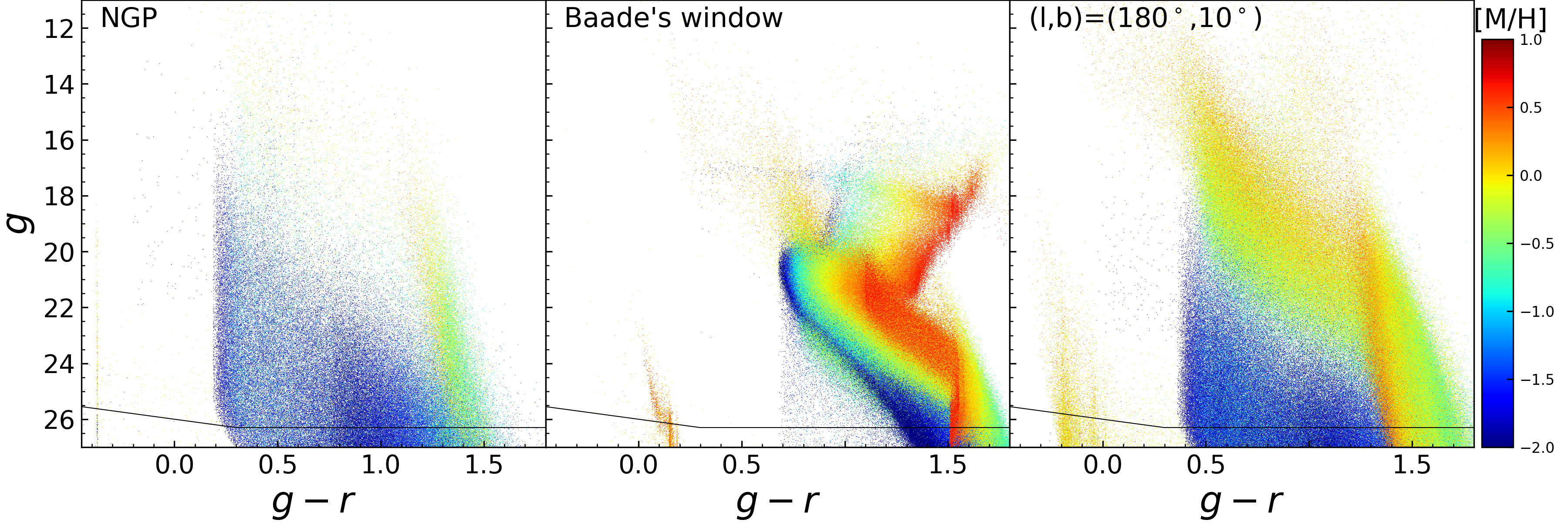}
\\
\includegraphics[width=0.78\textwidth]{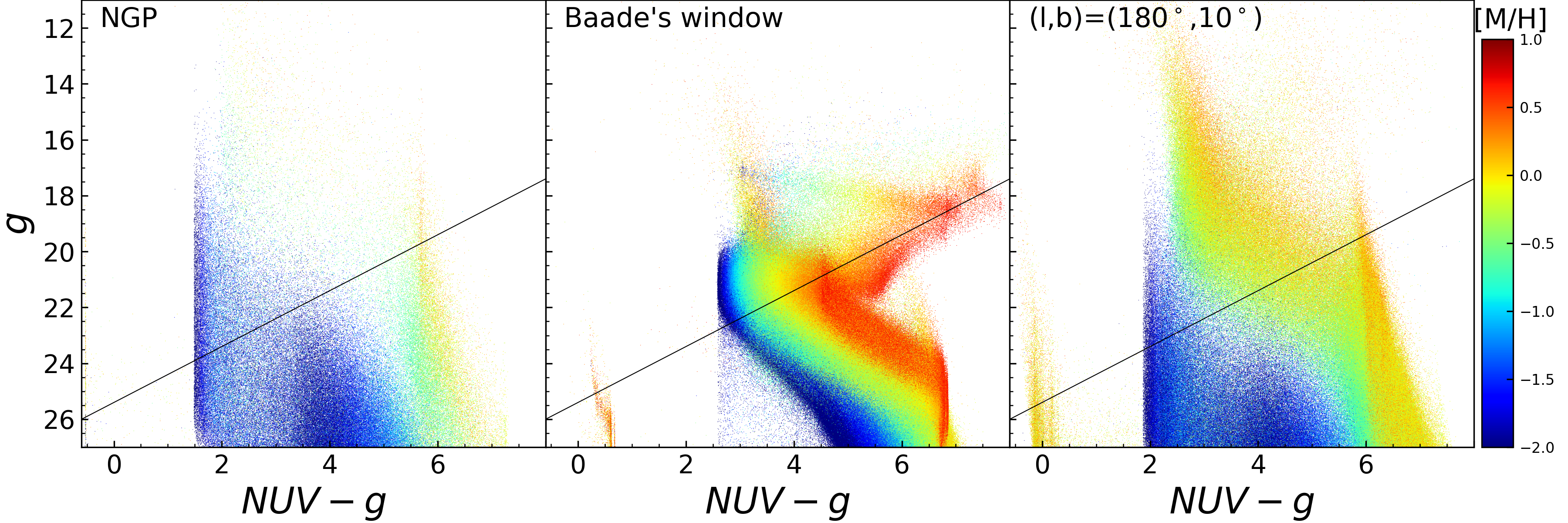}
\caption{$g$ vs. $g-r$ (upper panels) and $g$ vs. ${\rm NUV}-g$ (lower panels) for all stars within $2^\circ$ of the North Galactic Pole (NGP, left panels), Baade window (middle panel) and the anti-Galactic center direction (right panel). The black line in each panel indicates the magnitude limit of CSST-OS. Stars are color-coded by metallicity.}
\label{fig:cmd}
\end{figure*}

\subsection{Magnitudes and colors of the mock catalogue}

In this section, we present  illustrative cases  of the synthetic CMD and color-color diagrams from the mock catalogue. 
These plots are widely used for stellar population diagnostics including stellar metallicity determination \citep[e.g.][]{Hayden2015} and young star identification \citep[e.g.][]{Fang2017}. 
We have selected three typical positions on the sky to demonstrate the use of these plots: 
the North Galactic Pole (NGP), the Baade's window and the anti-Galactic center direction ($b=10^\circ$ and $l=180^\circ$).

Figure~\ref{fig:cmd} displays the CMD for stars in the three selected sky directions.
The upper panels shows CMDs with  $g-r$ versus $g$, while the lower ones are with  ${\rm NUV} -g$ versus $g$,  color-coded by metallicity.
The morphology of these diagrams differs due to the variation in the stellar populations across the three fields.
To demonstrate the observation coverage of CSST, we mark the CSST-OS magnitude limit in all panels of Figure~\ref{fig:cmd}. The lower panels clearly demonstrate that the NUV band,  regardless of the observation window, is useful only for identifying relatively hot and bright stars.

\begin{figure*}
\centering
\includegraphics[scale=0.68]{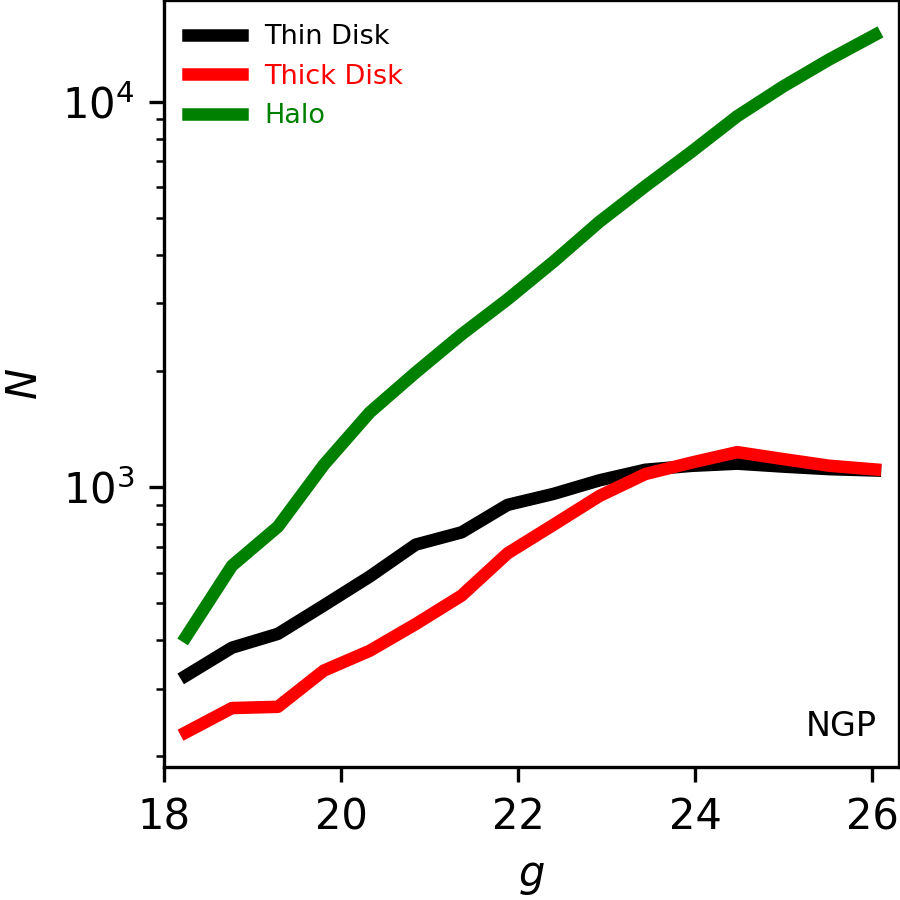}~
\includegraphics[scale=0.68]{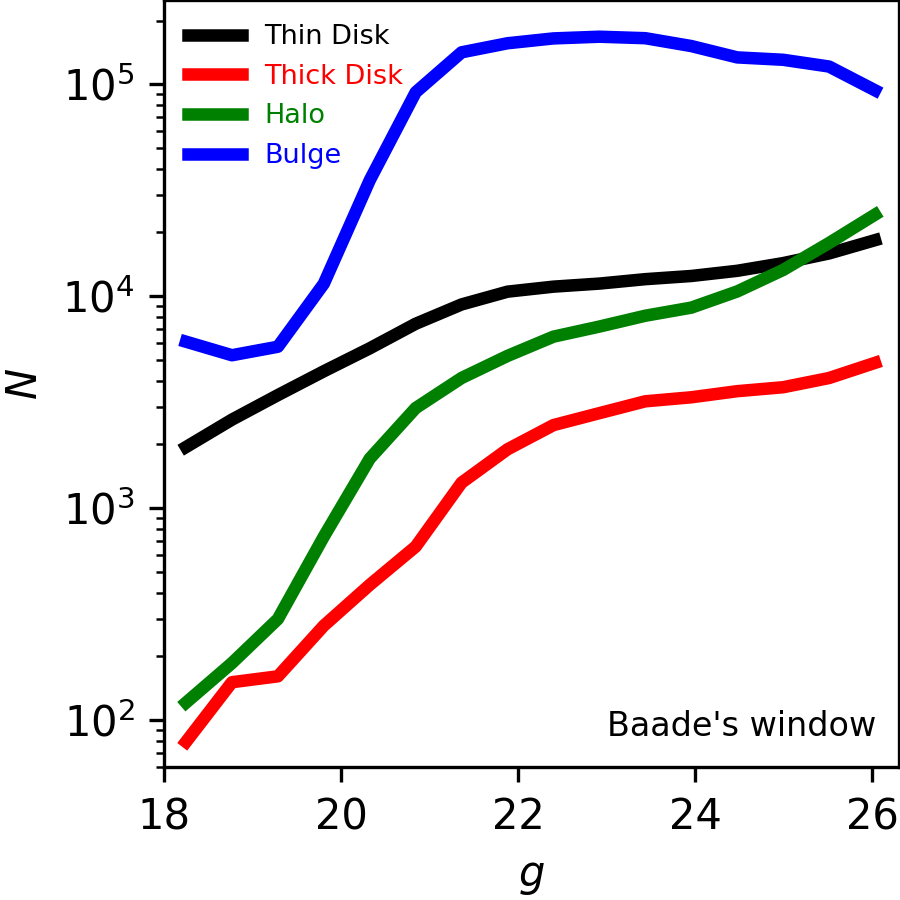}~
\includegraphics[scale=0.68]{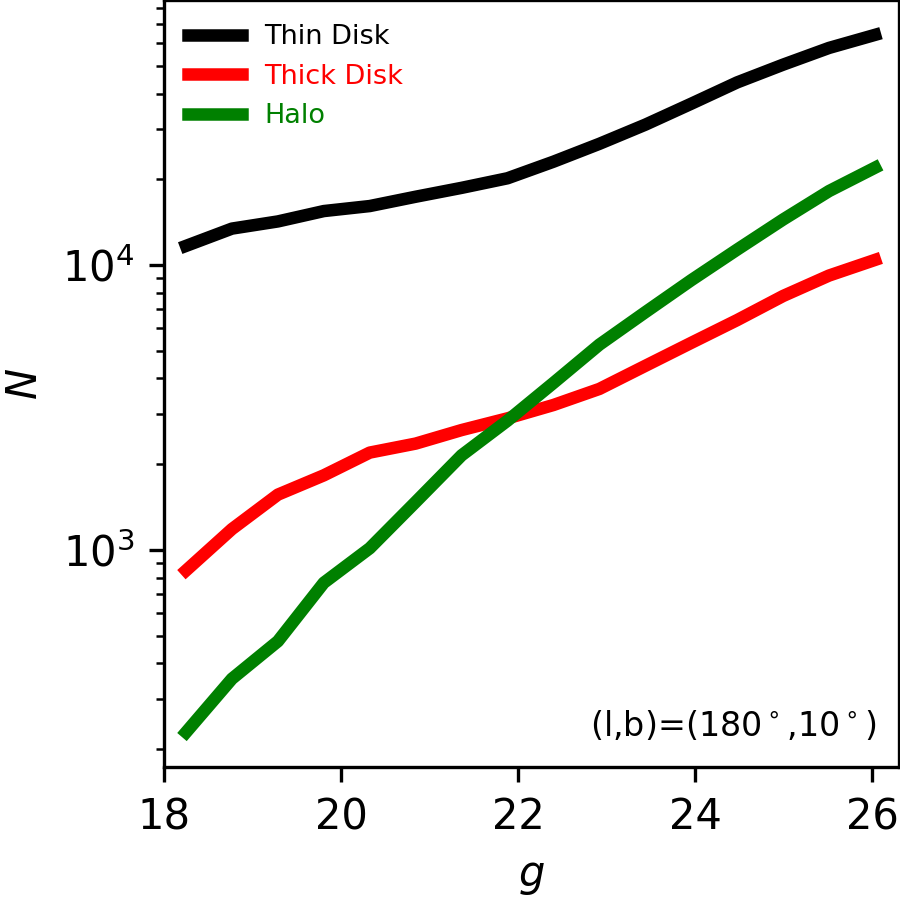}
\caption{The apparent $g$ band magnitude distribution function for the three locations shown in figures~\ref{fig:cmd}.}
\label{fig:g_mag_dist}
\end{figure*}

To illustrate the stellar contribution within the three selected fields, Figure~\ref{fig:g_mag_dist} represents the distribution function of $g$ band magnitudes for different galactic components.

\begin{figure*}
\centering
\includegraphics[scale=0.55]{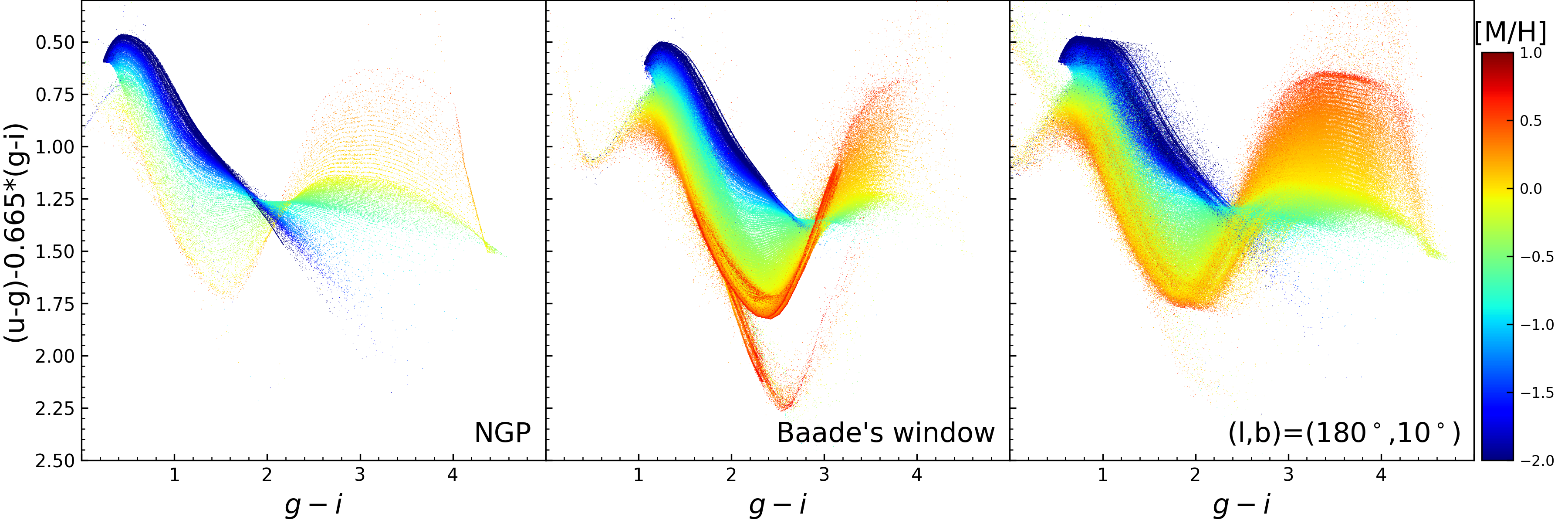}
\caption{Metallicity-sensitive color-color plot for the three locations shown in figures~\ref{fig:cmd}.}
\label{fig:uggi}
\end{figure*}

In figure~\ref{fig:uggi}, we show the $g-i$ versus $(u-g)-0.665*(g-i)$ daigram for the same locations as in figure~\ref{fig:cmd}, which shows well the dependence on metallicity. The coefficient of 0.665 is computed by taking the color excess ratio $E_{u-g}/E_{g-i}$ (with $A_u=1.553$, $A_g=1.197$ and $A_i=0.662$). As also shown in Chiti et al. (2021)  \cite{Chiti21}.
This plot can \checkit{be useful to select} and identify stars with different metallicities.

\begin{figure*}
\centering
\includegraphics[scale=0.55]{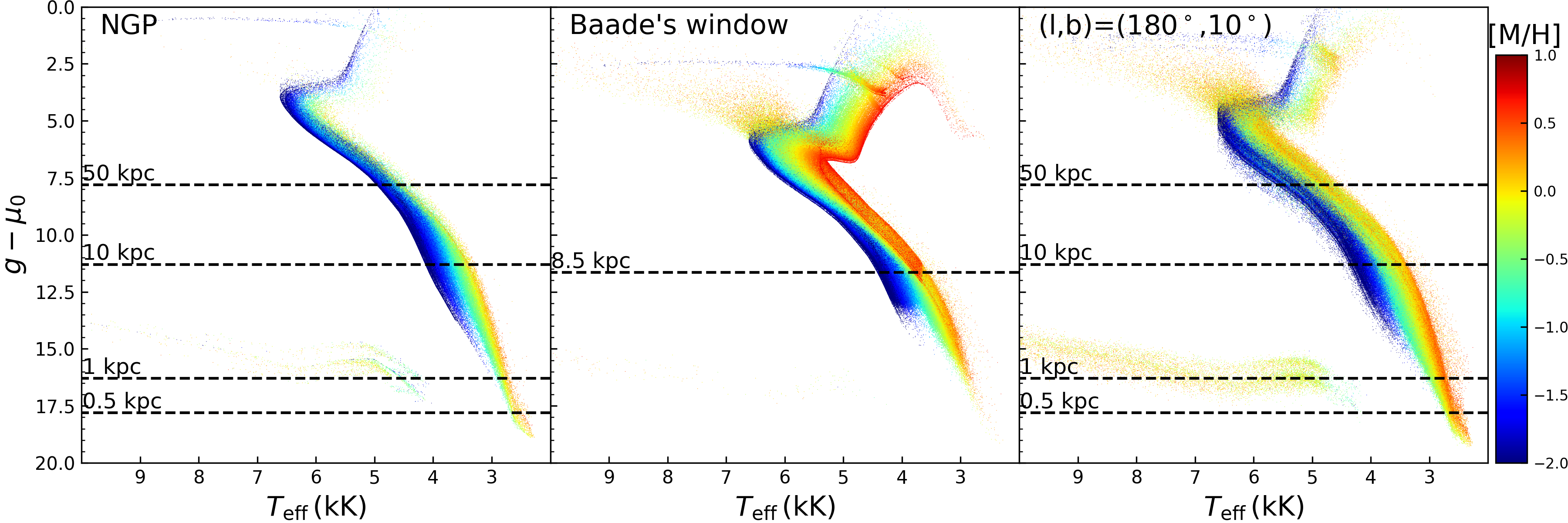}
\caption{$g-\mu_0$ vs. ${(T_{\rm eff})}$ for the three above locations, color coded with $\mh$. The horizontal lines indicates the limit of $g=26.3$\,mag for stars at different distances.}
\label{fig:mgteff}
\end{figure*}

In figure~\ref{fig:mgteff}, we presnet the $g$ band absolute magnitude versus the effective temperature for the three aforementioned locations. This figure reveals that very cool white dwarfs can be only observable in the Solar Neighbourhood (see the left and right panel).
Very low mass stars can be detected within a distance of less than 10\,kpc. 
For distance beyond  $\sim$ 50\,kpc, one can still use subgiants and brighter stars for relevant studies. 
Therefore, while the faint magnitude limit by CSST-OS will enable us to unlock the hidden nature that contributed by faint stars, particularly those at the outskirt of the Galaxy and the low mass end of stellar populations. Our mock data catalogue serves as a benchmark model for comparison with future real data.

\subsection{Kinematics}

The CSST-OS will provide proper motion measurement to the accuracy of better than 1 mas/yr. Combined with existing radial velocity measurement from \gaia  \cite{Seabroke2021}, it will depict the kinematic structure of the MW. 
Therefore, we provide the kinematic information for each star. First, we model the velocities of the stars with respect to the Galactic
center. Depending on the Galactic components, the stellar velocities are modelled with different velocity ellipsoids. For the thin disk stars, we take the velocity dispersion--age relation derived by Holmberg et al. (2009)  \cite{Holmberg09}. 
For the thick disk and halo stars, the velocity dispersion relation is from Chiba \& Beers (2000)  \cite{Chiba00}. 
For the bulge stars, we take the velocity dispersion values from Robin et al. (2003)  \cite{Robin03}.
These velocity ellipsoids are considered to rotate with respect to the local standard of rest (L.S.R., 220 $\rm km/s$), using values from Robin et al. (2003)  \cite{Robin03} 
with small adjustments. The velocity in the Galactocentric reference frame $\vector(V_x,V_y,V_z)$ is then transformed to the heliocentric reference frame velocity $\vector(U,V,W)$ with:
\begin{equation}
\displaystyle\vector(U,V,W)=\displaystyle\vector(V_x,V_y,V_z)-\displaystyle\vector(V_{x,\odot},V_{y,\odot},V_{z,\odot}),
\end{equation}
where 
[$V_{x,\odot},V_{y,\odot},V_{z,\odot}$] = [-10,5.25,7.17] $\rm km/s$ 
according to Dehnen \& Binney (1998)  \cite{Dehnen98}.
Although the derived velocity dispersion distribution may not obey the Jeans equation which is required for collisionless dynamic systems, they should give a decent comparison with the observations since they are derived from empirical relations.

\subsection{Stellar crowding analysis}

Observations of fields with high stellar surface density require a careful consideration of the crowding limit, regardless of the accuracy of photometry or spectroscopy.
In dense regions, e.g. galaxy central regions or the core of star clusters, light from multiple stars in the same field of view overlaps and blends together, making it difficult to separate and accurately measure the brightness of individual stars.
This effect is even more pronounced for fainter stars, as they are more numerous than the brighter ones. 
Indeed, the completeness of large photometric surveys such as \gaia is effectively limited by crowding rather than by the actual spatial resolution of the telescope  \cite{Cantat-Gaudin2023}.
For the sake of the observation accuracy and completeness, it is crucial to 
have a rigorous knowledge about the initial brightness and density of all sources in the field.

 The crowding and its corresponding photometric errors depend not only on the stellar density but also on the magnitude distribution of the stars  \cite{Olsen03,Cantat-Gaudin2023}. 
 Generally, evaluating the photometric error brought by crowding can be done with simulations, where  artificial stars are distributed across the CCD and the photometric errors are computed by comparing the artificial input magnitudes to the output ones as determined by the photometric data reduction pipeline. By varying the stellar density, the effect of crowding to the photometric error is then evaluated. 
 However, we adopt the approach of Olsen et al. (2003)  \cite{Olsen03} 
 that is based on an analytical model and is substantially less demanding in terms of time and computational resources.
 This model allows us to estimate the photometric error due to crowding for a given star with specified telescope resolution and stellar luminosity function. With the computed photometric errors, we can then obtain the so-called ``crowding limit''. This parameter is the magnitude limit at which a threshold photometric error $\sigma_{\rm th}$ is reached. 
 The stellar luminosity function, i.e. the magnitude distribution of the stars,  then influences the crowding limit results. 
 
 Different Galactic components show quite different magnitude distributions, as shown in figure~\ref{fig:g_mag_dist}.
 Thus we use the apparent magnitude to describe the crowding limit. 
 This choice also makes the simulation more directly connected to observations.

\begin{figure*}
\centering
\includegraphics[trim={0cm 0cm 0cm 0.15cm}, clip, scale=0.36]{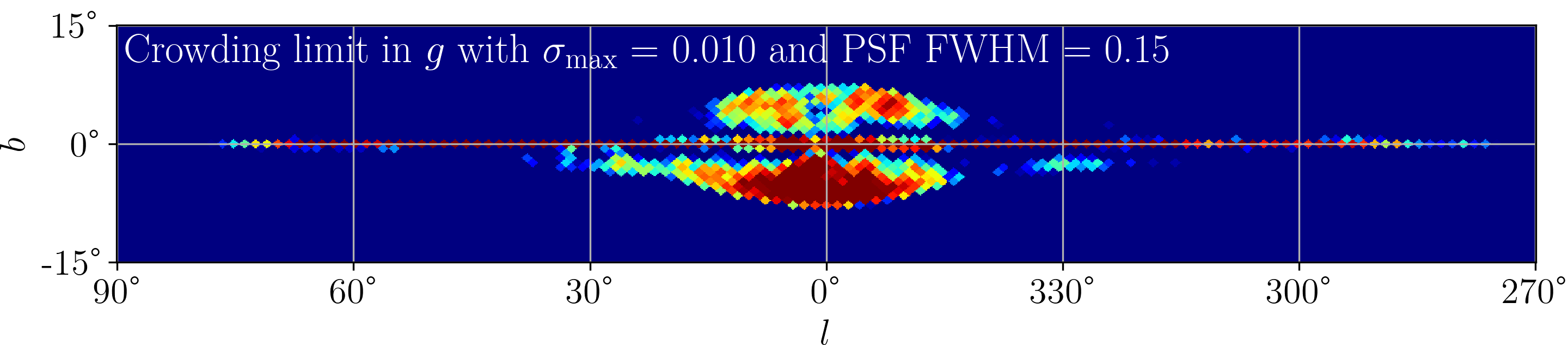}~
\includegraphics[trim={0cm 0cm 0cm 0.15cm}, clip, scale=0.36]{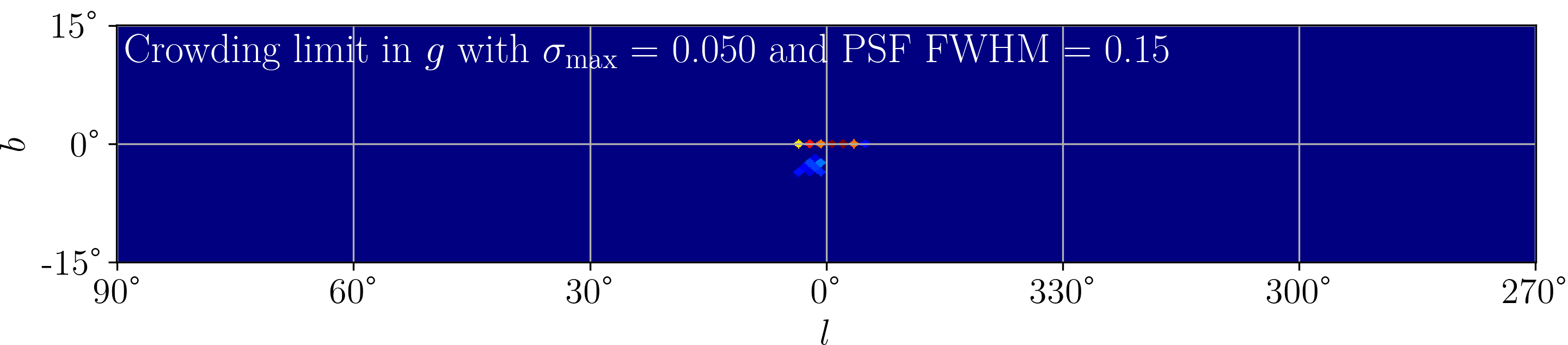}\\
\includegraphics[trim={0cm 0cm 0cm 0.15cm}, clip, scale=0.36]{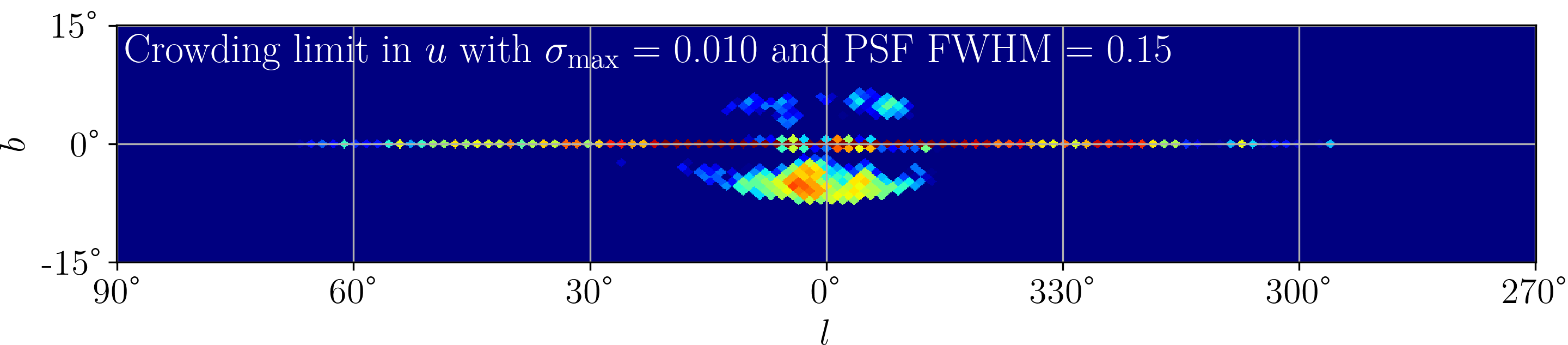}~
\includegraphics[trim={0cm 0cm 0cm 0.15cm}, clip, scale=0.36]{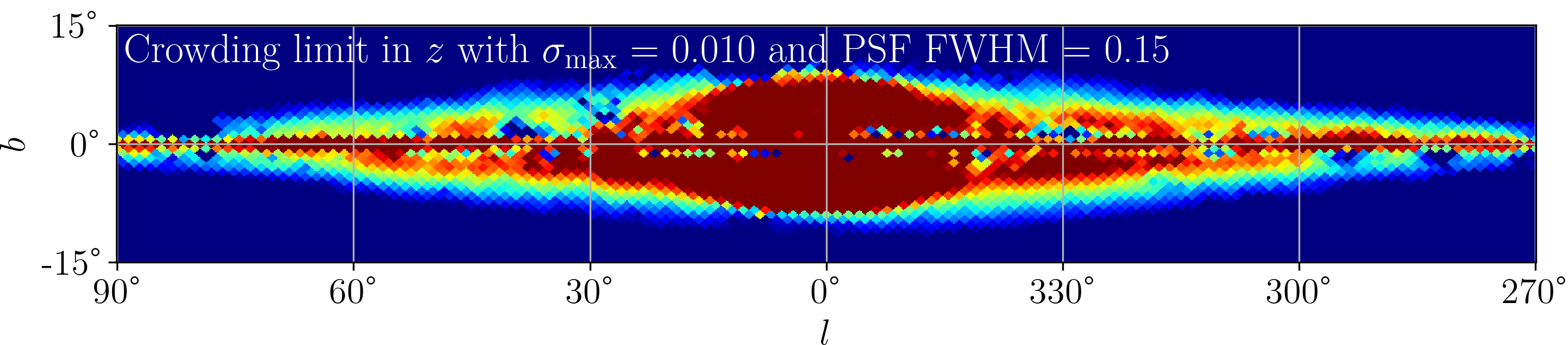}\\
\vspace{-.15cm}
\includegraphics[trim={6cm 0cm 6cm 4.3cm}, clip, scale=0.5]{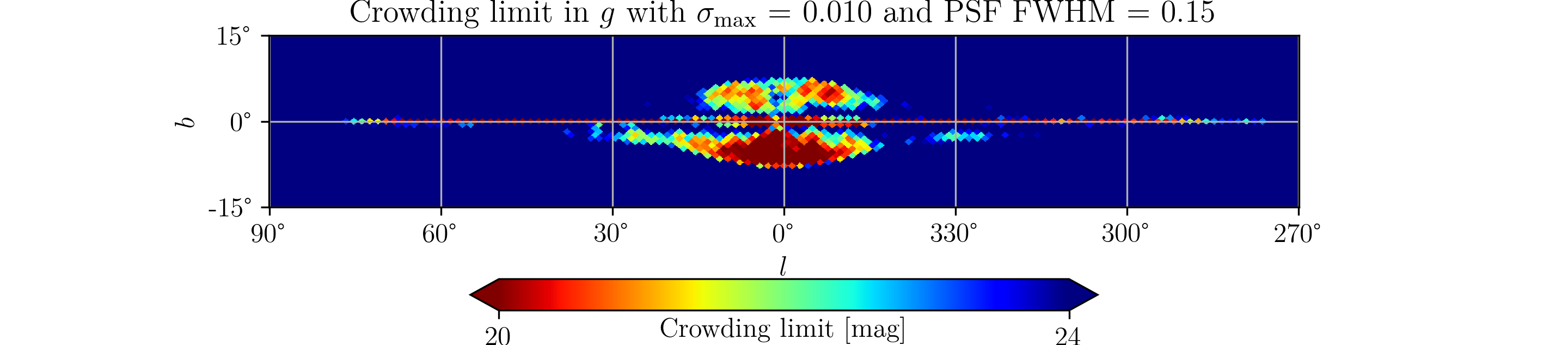}
\caption{Upper panels: crowding limit in the $g$ band at PSF FWHM 0.15$^{\dprime}$, with photometric errors of $\sigma=0.01$ (left panel) and $\sigma=0.05$ (right panel). Lower panels: crowding limit in the $u$ (left panel) and $z$ (right panel) bands at PSF FWHM of 0.15$\dprime$ and photometric error of $\sigma=0.01$.}
\label{fig:crowdlimit}
\end{figure*}

To calculate the crowding limit for CSST, the instrument PSF needs to be considered as well as the initial star number density and luminosity function.
We present the stellar crowding analysis following the method of Olsen et al. (2003)  \cite{Olsen03}. 
First, we compute the luminosity functions for each pixel. At each magnitude bin of the luminosity function, we integrate the flux of all stars within fainter magnitude bins and within a circled area with a diameter of the PSF FWHM.
Then this flux is taken as the photometric error for this magnitude. Once this error equals $\sigma_{\rm th}$, the magnitude is taken as the crowding limit. 
In this way, we construct the crowding limit map as shown in figure~\ref{fig:crowdlimit}, \checkit{where} the CSST $g$, $u$ and $z$ bands are shown as examples. 
For a given stellar density, a better PSF FWHM will result in a better photometric accuracy for fainter stars. 
The PSF FWHM adopted here, $\sim0.15\arcsec$, is the FWHM of the CSST PSF expected to be achieved  \cite{Zhan18}.
With this PSF, a photometric error $\sigma_{\rm max}=0.01$\,mag can be reached for most of the sky with $g>$ 24\,mag except for the bulge and low latitude regions. 
If $\sigma_{\rm max}=0.05$\,mag is taken as the acceptable photometric error, $g=24$\,mag can be achieved even for the bulge except for a few heavily attenuated regions. 
Therefore, we can conclude that it would be very valuable to include the low latitude regions in the CSST-OS survey. This is impossible to achieve for ground based telescopes because of their larger PSF FWHM and limited inclination. 
This is also impossible for HST because of its small FOV. 
CSST thus plays a unique role to help us understand the structure and origin of  these crowded regions.

We also note that the crowding issue becomes more severe towards longer wavelengths (i.e., from the $u$ band to the $z$ band)
though the settings of $\sigma_{\rm th}$ and PSF FWHM are the same. This mainly reflects the change of magnitude distribution of the stars as a function of wavelength. Considering worse PSF toward longer wavelength, the real crowding limit is brighter in the redder bands.

In summary, based on our crowding limit analysis, we demonstrate that the unique capabilities of CSST indeed allow a very deep and wide MW disk survey, even to fields near the Galactic plane. It could be a good reference for consideration to the CSST survey strategy design.

\subsection{Spectral interpolation routine for the mock catalogue}
Besides the imager, the CSST SC is also equipped with a slitless spectrograph in GU (255–420 nm), GV (400–650 nm), and GI (620–1000 nm). To facilitate the research concerning this device, we provide an interpolation routine which works on the mock catalogue presented in the current work. By reading the stellar physical parameters from the mock catalogue, it generates the corresponding model spectra for each star. Currently, it works with the \checkit{\phoenix\checkit{\footnote{\url{https://phoenix.ens-lyon.fr/Grids/BT-Settl/AGSS2009/SPECTRA/}}}}~\cite{phoenix} and \checkit{\CKmodel\checkit{\footnote{\url{https://wwwuser.oats.inaf.it/castelli/grids.html}}}}~\cite{ck03} atmosphere models, which is sufficient for the CSST spectrograph calibration work at the current stage, while other atmosphere models can be provided upon request.

\section{Simulations of galactic star clusters} \label{sec:cluster}

Star clusters provide ideal opportunities to study the evolution of simple stellar populations due to the minimal distance variance among their member stars. 
They also serve as excellent targets for testing the photometric quality of observations, particularly in crowded fields. 
In this section, we examine several simulated clusters with varying ages and metallicities to demonstrate the capabilities of the CSST mock catalogue.

\begin{figure}[H]
\centering
\includegraphics[scale=0.6]{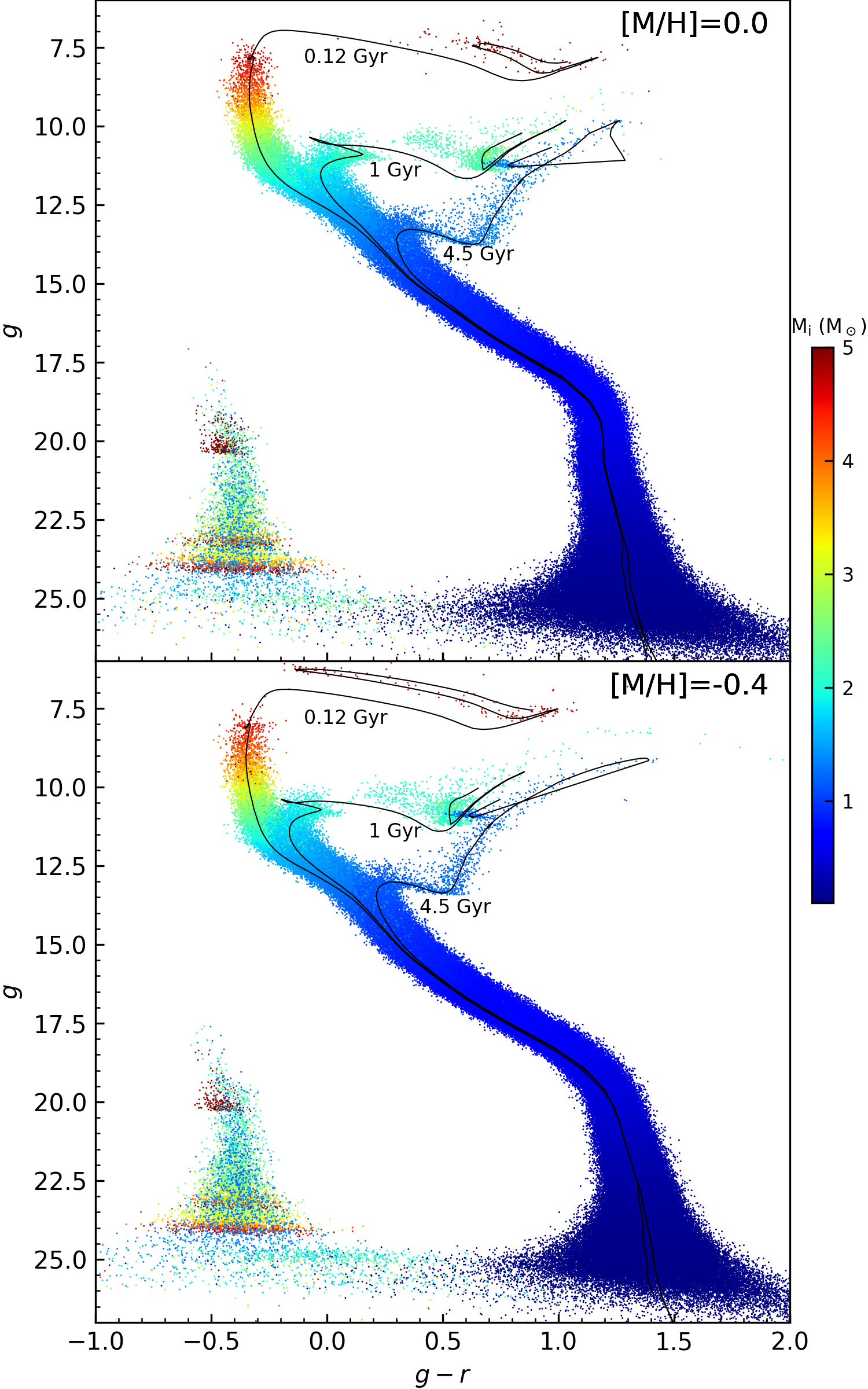}
\caption{CMD for simulated star clusters with Solar abundance (upper panel) and low-metallicity ($\mh=-0.4$, lower panel) and ages of 0.12\,Gyr, 1\,Gyr and 4.5\,Gyr respectively. The stars are placed 1\,kpc away.}
\label{fig:rich_cluster}
\end{figure}

In figure~\ref{fig:rich_cluster}, we show six simulated star clusters placed at 1\,kpc away. The clusters have different ages and metallicities, specifically, we consider three age values: 0.12\,Gyr, 1\,Gyr and 4.5\,Gyr, combined with two metallicities: Solar value and $\mh=-0.4$\,dex. 
All clusters are simulated with an initial stellar mass of $5\times10^5$ solar mass. 
The simulations take into account a photometric binary fraction of 0.3 and use the IMF from Chabrier et al. (2003)  \cite{Chabrier03}. 
No spread in age or metallicty is considered.

We add photometric error to the simulation with an approach similar to that in Wehner et al. (2008)  \cite{Wehner08}. 
We model the photometric uncertainties with:
\begin{equation}
\label{eq:sigma}
    \sigma_{\rm m}=0.01+0.2e^{m-m_{\rm lim}}
\end{equation}
and the completeness with the Pritchet function  \cite{Fleming1995}:
\begin{equation}
    f=\frac{1-\alpha(m-m_{\rm lim})}{2\sqrt{1+\alpha^2*(m-m_{\rm lim})^2}},
\end{equation}
where $m$ is the magnitude of a given star in a given band. 
According to the CSST white paper, we adopt $m_{\rm lim}={\rm %
[25.4,25.4,26.3,26.0,25.9,25.2,24.4]}$\checkit{.}
\checkit{For $\alpha$,} we take the values from Wehner et al. (2008), $\alpha={\rm [0.8,0.8,0.8,1.4,1.4,1.4,1.4]}$, to give some qualitative indications.
For example, by taking $g=$26.3 and 20\,mag, they give $\sigma_{\rm m}=$0.21 and 0.01, respectively.

The star cluster mock catalogue serve as a valuable resource, empowering researchers to plan and conduct star cluster related studies with CSST.
Figure \ref{fig:rich_cluster} demonstrates CSST's potential to provide high quality data reaching as faint as $g\sim24$\,mag for star clusters at a distance of 1\,kpc. 
This capability allows for the detection of the white dwarf cooling sequence in nearby clusters. 
Additionally, the placement of simulated star clusters at various distances within our Galaxy can aid researchers in studies such as identifying cluster members, deriving cluster parameters (e.g. differential reddening analysis), and potentially discovering new star clusters using real CSST data.

\section{Summary} \label{sec:sum}
In this study, we  present a comprehensive mock stellar catalogue of the Milky Way using the \trilegal~ stellar population synthesis tool. 
The catalogue is specifically designed for the CSST SC photometric system and encompasses the main stellar components of the Galaxy, including the thin disk, thick disk, halo, and bulge. It consists of approximately 12.6 billion stars, reaching magnitudes as faint as 27.5 mag.
For each star in the catalogue, we provide a wide range of information, including stellar physical parameters (the effective temperature, surface gravity, luminosity, age and \checkit{surface metallicity}), photometric information %
, astrometry (true distance modulus, right ascension and declination), as well as kinematics (Galactic spatial velocity, radial velocity and proper motion). 
The generated mock stellar catalogue is available for retrieval from the VO database. 

In addition to the main stellar catalogue, we offer an interpolation routine that generates spectral energy distributions based on the provided catalogue.
This feature is particularly  valuable for researchers involved in slitless spectroscopic surveys. 

We also provide catalogues specifically dedicated to  star clusters, encompassing a range of ages and metallicities. These catalogues serve as valuable resources for studies of stellar cluster evolution and crowded-field stellar photometry.

The creation of this catalogue represents a milestone for the CSST, as it is the first of its kind specifically tailored for the mission. Its availability will contribute to the development of CSST by providing a more realistic input for survey strategy planning, photometric system calibration, and validation of the data reduction pipeline
The catalogue holds immense potential to advance a wide range of research studies relevant to the CSST. It can serve as a valuable resource for investigations in stellar physics, Milky Way structure, and extragalactic studies. Researchers can utilize the catalogue to train algorithms for the selection of different types of stars, explore the dependence of stellar properties on parameters such as metallicity and age, and analyze variations across different regions of the Galaxy.

By conducting crowding limit analysis, we propose extending the CSST-OS survey to encompass low Galactic latitude regions. This extension would enable the exploitation of CSST's unique capabilities in regions where stellar crowding presents challenges. Such an expansion would unlock new opportunities for scientific exploration and maximize the scientific output of the CSST mission.

This catalogue serves as a  representation of our current understanding of the main stellar populations in the MW. 
However, we anticipate that a more updated and refined picture of the MW will emerge once \checkit{CSST} completes its Optical Survey, benefiting from its exceptional survey depth.
The \checkit{CSST} will unveil the bulk of faint stars that have remained hidden from previous optical sky surveys. This will allow for a direct comparison between the observed data and the information provided in our catalogue. 
Such a comparison has the potential to shed light on our knowledge of low-mass stars and distant stars within the Milky Way. It will also facilitate the identification and analysis of sub-structures present within our \checkit{Galaxy}.

\vspace{0.3cm}
{\noindent
\scriptsize %
\it 
This work is supported in part by the National Key R\&D Program of China (2021YFC2203100, 2021YFC2203104) and the science research grants from the China Manned Space Project with NO. CMS-CSST-2021-A08.
Y.C. acknowledges  National Natural Science Foundation of China (NSFC) No. 12003001 and the Anhui Project (Z010118169). 
X.F. thanks the support of the NSFC No. 12203100. 
D.F. acknowledges NSFC No. 12273077. \checkit{P.M.  acknowledges support from Padova University through the research project PRD 2021.} 
We thank  Xianmin Meng for providing the CSST filter transmission curves, thank Di Li for valuable suggestions, China National Astronomical Data Center (NADC) and Chinese Virtual Observatory (China-VO) for hosting our catalogue. \checkit{We acknowledge the anonymous reviewers for useful suggestions.}
}

\vspace{0.3cm}
{\noindent \footnotesize
{\bf Conflict of interest}\quad The authors declare that they have no conflict of interest.}

{
\footnotesize
\bibliography{ref}
\bibliographystyle{aasjournal}
}

\end{multicols}

\begin{appendix}
\renewcommand{\thesection}{Appendix}
\section{}

\subsection{Comparison of future wide-field optical space telescopes}

\begin{table}[H]
\centering
\footnotesize
\caption{Comparison of the CSST SC with the imaging instruments of the Euclid and Roman telescopes.}
\label{tab:telescope}
\scriptsize
\begin{tabular*}{0.87\textwidth}{l|l|l|l} \hline
  & CSST/SC & Euclid/VIS/NISP & Roman/WFI \\ \hline
 Orbit       & low earth orbit & L2 & L2 \\ \hline
 Field of View (deg$^2$)       & 1.10 & 0.55 & 0.281 \\ \hline
 PSF FWHM ($\dprime$) & 0.15 & 0.16  & 0.127 \\ \hline
 Primary mirror diameter (m)  & 2    & 1.2  & 2.4 \\\hline
 Wavelength range (nm) & 255-1000 & 550–2000 & 480-2000 \\ \hline
 Wide survey area (deg$^2)$ and depth (mag) & 17,500, $\sim26$ & 15,000, $\sim27$ in VIS, $\sim24$ in NISP & 170, $\sim26.9$ \\ \hline
 Deep survey area (deg$^2)$ and depth (mag) & 400, $\sim 27$ & 40, $\sim$2\,mag. deeper than the wide survey & 40,  $\sim29$ \\ \hline
 Planned launch time & $\sim$2024 & July 2023 & Oct. 2025 \\ \hline
 Ref. &  \cite{Zhan18,Zhan21} & ~ \cite{Euclid} &  \cite{WFIRST,Rubin2021} \\ \hline
\end{tabular*}
\end{table}

\subsection{Milky Way  extinction map comparison}
\label{app:dust}
\begin{figure}[H] %
\centering
\includegraphics[scale=0.396]{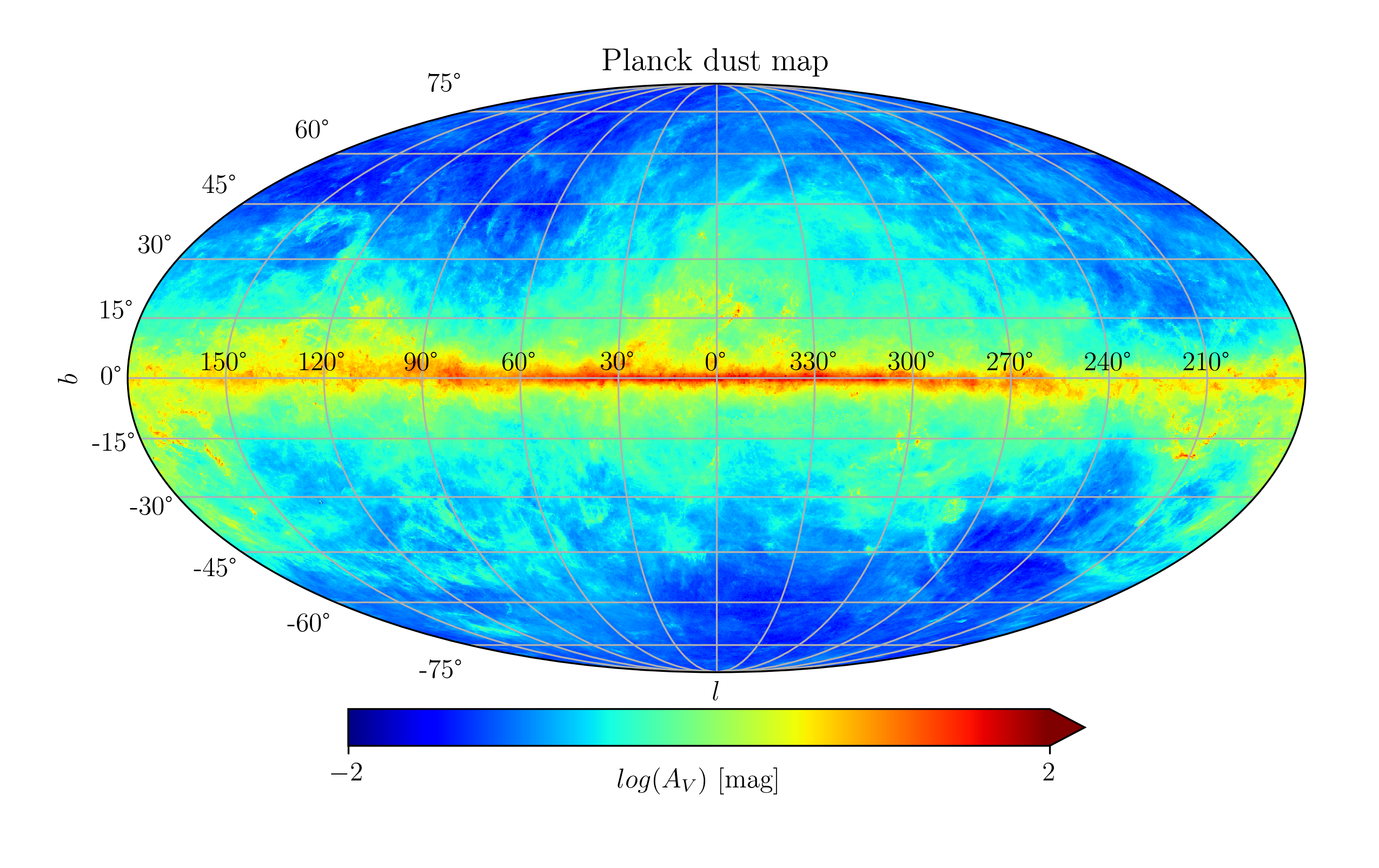}
\includegraphics[scale=0.396]{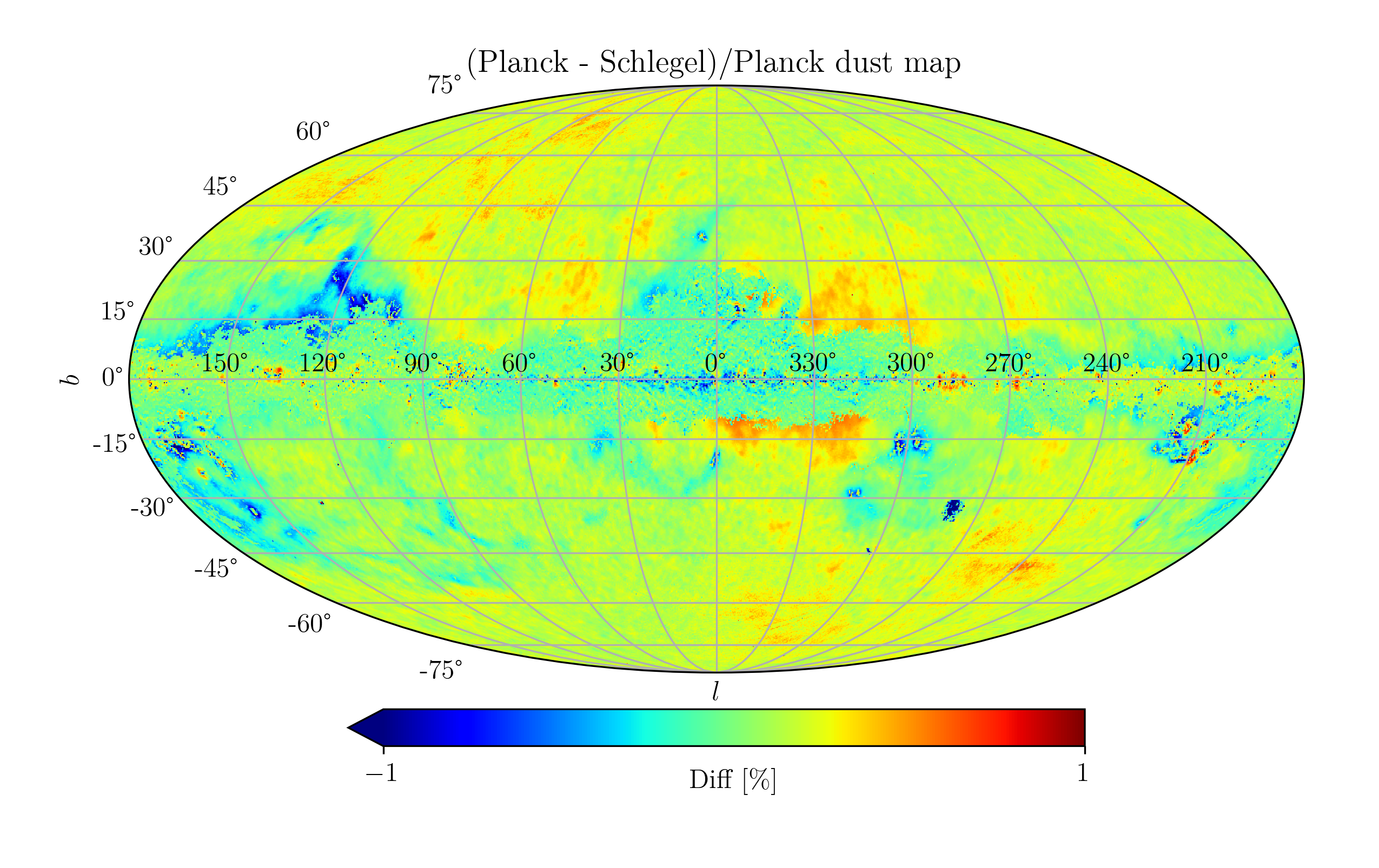}
\caption{Planck mean extinction map (upper panel)  \cite{Planck-dust} 
and the fractional difference of the Schlegel map  \cite{Schlegel} compared to the Planck map (lower panel). Both are displayed with $\nside=512$.
}
\label{fig:avmaps}
\end{figure}

In Figure~\ref{fig:avmaps} we present the Planck mean extinction map as well as the difference compared to the Schlegel one \cite{Schlegel} (with $\nside=512$, which is the highest resolution available for the Schlegel one). %

\subsection{Mock catalogue description}
\label{app:table_description}
In table~\ref{tab:cat}, we present the sample section of an example fits file. The meaning of the columns are:
\begin{description}
\item[Gc] Galactic components: ``1" for the thin disk,``2" for the thick disk, ``3" for the halo, ``4" for the bulge, and for ``5" optional objects located at fixed distance (in case of star clusters or external galaxies).
\item[logAge] logarithm of the age in years.
\item[M\_H] initial metal abundance, $\rm [M/H]$. %
\item[m\_ini] initial stellar mass in unit of $\msun$.
\item[mu0] true distance modulus $\mu_0$ in unit of magnitude.
\item[Av] extinction in the $V$ band, the Cardelli et al. (1989)  \cite{Cardelli89} 
 extinction curve with $R_V=3.1$ is used in this simulation.
\item[mratio] mass ratio of binaries, 0.0 for single stars, value between 0.0 and 1.0 for primaries.
\item[Mass] the current stellar mass in unit of $\msun$.
\item[logL] logarithmic luminosity in solar unit, $\log(L/L_\odot)$.
\item[logTe] logarithmic effective temperature, $\log(T_\mathrm{eff}/\mathrm{K})$.
\item[logg] $\log$ of surface gravity in cm/s$^2$.
\item[label] evolutionary stage. 0 for pre-main sequence (PMS), 1 for main sequence (MS), 2 for subgiant branch (SGB), 3 for red giant branch (RGB), 4 for the initial red part of core helium burning (CHeB) phase, 5 for the blueward loop of CHeB phase, 6 for the redward loop of CHeB phase, 7 for early-AGB, 8 for TP-AGB, 9 for post-AGB and white dwarf (WD) phase.
\item[McoreTP] core mass in $\msun$ (valid only for TP-AGB and successive evolutionary stages).
\item[C\_O] C/O, surface carbon and oxygen abundance ratio (in number).
\item[period0, period1] expected LPV fundamental mode and first overtone periods in days.
\item[pmode] expected dominant period (between the two above).
\item[Mloss] mass loss rate in $M_\odot/\mathrm{yr}$.
\item[tau1m] optical depth of circumstellar dust at 1$\mu$m.
\item[X, Y, Xc, Xn, Xo] surface mass fractions of hydrogen, helium, carbon, nitrogen, and oxygen.
\item[Cexcess] carbon excess, defined as ${\rm log}_{10}(n_{\rm C} - n_{\rm O} ) - {\rm log}_{10} ( n_{\rm H} ) + 12 $, with $n_{\rm C}$, $n_{\rm O}$, and $n_{\rm H}$ being the surface number fractions of carbon, oxygen, and hydrogen, respectively.
\item[Z] surface mass fraction of metals.
\item[mbolmag] bolometric magnitude.
\item[NUVmag, umag, gmag, rmag, imag, zmag, ymag] apparent magnitudes in CSST photometric bands, in AB magnitude system.
\item[usmag, gsmag, rsmag, ismag, zsmag] apparent magnitudes in SDSS photometric bands, in AB magnitude system.

\item[gp1mag, rp1mag, ip1mag, zp1mag, yp1mag, wp1mag] apparent magnitudes in Pan-STARRS1 photometric bands, in AB magnitude system.

\item[velU, velV, velW] three-dimensional heliocentric velocities ($\rm U$, $\rm V$, $\rm W$ ) in km/s.
\item[Vrad] radial velocity in km/s.
\item[PMracosd, PMdec] proper motions $\mu_\alpha \cos\delta $ and $\mu_\delta$, in mas/yr.
\item[gall, galb] Galactic longitude and latitude in degrees.
\item[ra, dec] equatorial right ascension and declination in degrees.

\end{description}
We caution that the binary stars in this simulation are non-interacting binaries. The results with the code including the interacting binaries, developed in Dal Tio et al. (2021)  \cite{DalTio21}, 
will be presented in the next releases.

\begin{table*}
\footnotesize
\caption{Example of the mock catalogue.}
\label{tab:cat}
\begin{tabular*}{\textwidth}{cccccccccccccc}
\hline
\hline
Gc &logAge &M\_H &m\_ini &mu0 &Av &mratio &Mass &logL & logTe &logg &label &McoreTP &C\_O \\
\hline
  1 & 8.21 & 0.42 & 1.89 & 14.90 & 0.083 & 0.0 & 1.89 & 1.05 & 3.90 & 4.22 & 1 & 0.0 & 0.54 \\
  1 & 8.25 & 0.42 & 1.57 & 15.20 & 0.078 & 0.0 & 1.57 & 0.70 & 3.84 & 4.25 & 1 & 0.0 & 0.54 \\
\hline\hline
\end{tabular*}
\begin{tabular*}{\textwidth}{ccccccccccccc}
period0 &period1 &pmode &Mloss &tau1m & X &Y &Xc &Xn &Xo & Cexcess &Z & mbolmag \\
\hline
 0 & 0 & -1 & -1.00E-31 & 0 & 0.65 & 0.31 & 0.0063 & 0.0017 & 0.016 & -1 & 0.035 & 17.04 \\
 0 & 0 & -1 & -6.69E-14 & 0 & 0.65 & 0.31 & 0.0063 & 0.0017 & 0.016 & -1 & 0.036 & 18.21 \\
\hline\hline
\end{tabular*}
\begin{tabular*}{\textwidth}{cccccccccccccc}
NUVmag &umag &gmag &rmag &imag &zmag &ymag & usmag &gsmag &rsmag &ismag &zsmag & gp1mag & rp1mag  \\
\hline
 19.00 & 18.03 & 17.06 & 17.07 & 17.20 & 17.31 & 17.32 & 18.21 & 17.06 & 17.07 & 17.18 & 17.31 & 17.05 & 17.07 \\
 20.55 & 19.42 & 18.37 & 18.17 & 18.18 & 18.24 & 18.24 & 19.53 & 18.39 & 18.17 & 18.17 & 18.23 & 18.35 & 18.17 \\
\hline\hline
\end{tabular*}
\begin{tabular*}{\textwidth}{cccccccccccccc}
ip1mag & zp1mag &yp1mag &wp1mag &velU &velV &velW &Vrad &PMracos &PMdec &gall &galb &ra & dec\\
\hline
 17.19 & 17.30 & 17.32 & 17.10 & -19.68 & -62.68 & 8.53 & 19.88 & 0.69 & -1.27 & 204.89 & -0.21 & 98.98 & 7.06 \\
 18.17 & 18.23 & 18.24 & 18.22 & -59.55 & -47.37 & -18.51 & -22.72 & 0.17 & -1.40 & 204.82 & -0.12 & 99.04 & 7.16 \\
\hline
\end{tabular*}
\end{table*}

\end{appendix}

\end{document}